# Repulsion of polarized particles from two-dimensional materials

Francisco J. Rodríguez-Fortuño, Michela F. Picardi and Anatoly V. Zayats

*Department of Physics, King's College London, Strand, London WC2R 2LS, United Kingdom*

**Abstract** – Repulsion of nanoparticles, molecules and atoms from surfaces can have important applications in nanomechanical devices, microfluidics, optical manipulation and atom optics. Here, through the solution of a classical scattering problem, we show that a dipole source can experience a robust and strong repulsive force when its near-field interacts with a two-dimensional material that has a metallic character. As an example, the case of graphene is considered, showing that a broad bandwidth of repulsion can be obtained spanning the frequency range $0 < \hbar\omega < (5/3)\mu_c$, where $\mu_c$ is the chemical potential of graphene, tuneable electrically or by chemical doping.

Since the confirmation that light carries momentum in the early 20th century, the study of the mechanical force that light exerts on matter has developed into important scientific and technological applications[1], [2]. A simple example of optical force occurs when polarized particles are attracted to a nearby surface: any material surface brought close to an oscillating dipolar particle will experience oscillations of its constituent charges, whose scattered fields then exert forces back on the polarized particle. Under the quasistatic approximation, this is typically explained by an effective image dipole induced in the material[3]. This force can be very strong in the near field and is usually attractive for conventional materials: its influence is behind the unwanted adhesion and stiction in nanomechanical devices[4], [5]. Interestingly, recent works show that the surface material's optical properties can turn this attraction into a repulsion, even if the particle is in free space[6], [7]. The polarized particle can be any particle that exhibits a dipole-like electromagnetic field, ranging from small illuminated nanoparticles to single atoms. This optical repulsion of polarized particles from surfaces could lead to interesting novel applications, providing a simple route for levitation of particles or atoms away from a neighbouring surface by relying on the optical properties of a surface, instead of requiring structured illumination.

In order to achieve repulsion of a particle from a surface, the image dipole induced on the surface must oscillate in phase with the polarized particle, instead of out-of-phase as in conventional dielectrics: in previous works this was achieved by using materials whose relative electric permittivity is near zero[6]. These materials allow no electric field perpendicular to their boundary, intuitively squeezing the electric field between the dipole and the surface, resulting in repulsion[6]. This property, however, is available in natural materials only on narrow frequency bands, and is difficult to synthesize artificially[8], [9]. A viable alternative is to use anisotropic materials in which only one of the components of the permittivity tensor is near zero[7], retaining the repulsive behaviour. Although this condition can be achieved with artificial materials such as metal-dielectric stacks, the fabrication is still challenging, as finely adjusted thicknesses of metal and dielectric layers are needed to achieve the repulsion in a required spectral range.

In this work, we propose a much simpler approach to the repulsion of polarized particles from surfaces. Instead of relying on an optical substrate with bulk optical properties, we design the optical force acting on a polarized particle in the vicinity of a two-dimensional (2D) sheet, and we find a broad range of parameters that can result in repulsion of the polarized particle in a wide optical bandwidth. In Ref. [7] we found that thinner sheets of metal resulted in broader repulsion bands. This inspired us to investigate the thinnest materials, one-atom-thick two-dimensional materials, which are known to show metallic character under certain conditions. Two-dimensional materials have been lately shown to have truly remarkable properties, very different to bulk materials, and several practical examples of 2D materials such as graphene, transition metal dichalcogenides (TMDC), and boron nitride are being widely used in nanophotonics and optoelectronics[10]. We show that they can also exhibit large dipole repulsion bandwidths. The presented model also applies to topological insulators, whose surface can be regarded as a sheet conductivity. The model could explain the experimentally observed unusual wetting characteristics of graphene[11], [12] and Van der Waals forces on molecules near TMDCs[13], [14].

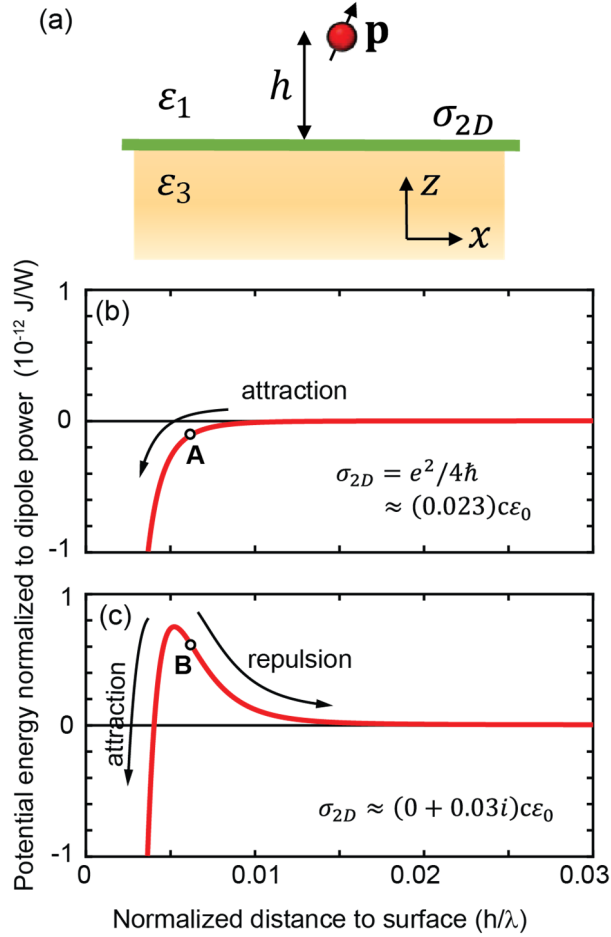

**Fig. 1**. (a) Geometry of the problem: a dipole **p** radiating at frequency $\omega$ (wavelength $\lambda$) located a distance $h$ above a 2D material surface on a substrate of relative permittivity $\varepsilon_3$. The upper medium has relative permittivity $\varepsilon_1$. (b,c) Numerically calculated potential energy landscape of a horizontal dipole as a function of its distance above the surface, for different values of surface conductivity equal to (b) ideal conductivity of graphene $\sigma_{2D} = e^2/4\hbar$ (corresponding to label "A" in Figs. 2-3), and (c) a two-dimensional conductivity with a high imaginary part, indicating metallic character (corresponding to label "B" in Figs. 1-3). The potential energy is normalized to the power radiated by the dipole $P_{\rm rad} = |\mathbf{p}|^2\omega^4/(12\pi\varepsilon_0\varepsilon_1(c_0^3/n_1^3))$, where $n_1 = \sqrt{\varepsilon_1}$.

**Optical force on dipoles above two dimensional sheets -** We consider a dipole source $\mathbf{p} = (p_x, p_y, p_z)$ radiating with a frequency $\omega$ at a position $\mathbf{r_0} = (0,0,h)$ above a two dimensional sheet conductivity $\sigma_{2D}$ at the plane $z = 0$, sandwiched between a superstrate and substrate with respective relative permittivities $\varepsilon_1$ and $\varepsilon_3$, as in Fig. 1(a). The time-averaged optical force acting on the dipole is given by[15]:

$$\langle \mathbf{F} \rangle = \sum_{i=x,y,z} \frac{1}{2} Re\{p_i^* \nabla E_i\} = \sum_{i,j=x,y,z} \frac{1}{2} Re\{p_i^* p_j \nabla G_{ij}\}$$

where $\nabla$ is the gradient with respect to $\mathbf{r}$ evaluated at $\mathbf{r_0}$, and $\mathbf{E} = (E_x, E_y, E_z)$ is the electric field acting on the dipole, given by $\mathbf{E}(\mathbf{r}) = \overleftrightarrow{\mathbf{G}}(\mathbf{r}, \mathbf{r_0}) \mathbf{p}$. The Green tensor $\overleftrightarrow{\mathbf{G}}$ accounts for the reflected field of a dipole above an arbitrary surface and is a function of the Fresnel reflection coefficients of the surface, involving transverse wave-vector integration[16], [17]. Although a lateral force may exist for circularly polarized dipoles[18]–[21], corresponding to the terms $\nabla G_{ij}$ with $i \neq j$, here we are interested on the vertical force component only, which is given by the terms $\nabla G_{ii}$ (see Supplementary Information) and evaluates to[6], [7]:

$$\langle F_z \rangle = \frac{1}{2} Re \left\{ \frac{-1}{8\pi\varepsilon_0\varepsilon_1} \int_0^\infty k_t \left[ \left(|p_x|^2 + |p_y|^2\right)(k_1^2 r^s - k_{z1}^2 r^p) + |p_z|^2(2k_t^2 r^p) \right] e^{ik_{z1}2h} \, dk_t \right\} \quad (1)$$

where $k_1 = k_0 n_1 = 2\pi n_1/\lambda_0$ is the wavevector in the upper medium with refractive index $n_1$ and $k_t$ is the transverse wavevector. The Fresnel reflection coefficients $r^p(k_t)$ and $r^s(k_t)$ of the two dimensional sheet (given in the Supplementary Information) can be substituted into Eq. 1, which can be numerically integrated. For linear dipoles, the optical force acts exclusively along z and $\langle F_z(h) \rangle$ is conservative, which allows us to calculate the potential energy landscape $U(h)$ of the dipole in the vicinity of the surface.

A remarkable phenomenon arises: for certain values $h \ll \lambda$ there can be a strong near-field repulsive force acting on the dipole, depending on $\sigma_{2D}$. This is the main result of this paper. When the imaginary part of $\sigma_{2D}$ is close to zero, the potential landscape of the dipole is such that the dipole will be strongly attracted to the surface (Fig 1(b)) as usual in most surfaces. However, when the values of $\sigma_{2D}$ have a positive imaginary part, associated with a metallic character of the 2D material enabling it to support surface waves, then the dipole will be strongly repelled away from the surface if it is beyond a certain threshold distance (Fig 1(c)). These values of $\sigma_{2D}$ are easily achieved in experimental two-dimensional materials as shown later. This behaviour is very different to previous works on dipole repulsion above surfaces, in which the force shows a simple $h^{-4}$ decay with distance in the near field under the quasistatic approximation, with no changes of sign in the force, except for periodic oscillations in the relatively weak far-field force caused by phase retardation[6], [7]. In the present case, the quasistatic limit $k_t \to \infty$ predicts an attractive force as $h \to 0$ but the force undergoes a change of sign in the near field, not associated with far-field phase retardation, which changes the force to being repulsive above a threshold distance. This suggests the presence of competing near-field phenomena. In Fig 2(a) we plot the dependence of the threshold distance of repulsion, i.e. the contour at which $\langle F_z(z) \rangle$ changes sign for different values of $h$, as a function of the complex conductivity of the surface $\sigma_{2D}$. These contours enclose the regions of $\sigma_{2D}$ where repulsion occurs for various distances. Fig. 2(b) plots the magnitude of the repulsive force at a fixed distance $h = 0.006\lambda$. From a macroscopic optics point of view, ignoring the atomic details, the only relevant optical parameter of a two-dimensional sheet is its complex sheet conductivity $\sigma_{2D}$ at a certain frequency. Thus, Fig. 2 provides a general recipe to the existence of repulsion from two-dimensional sheets at any given distance.

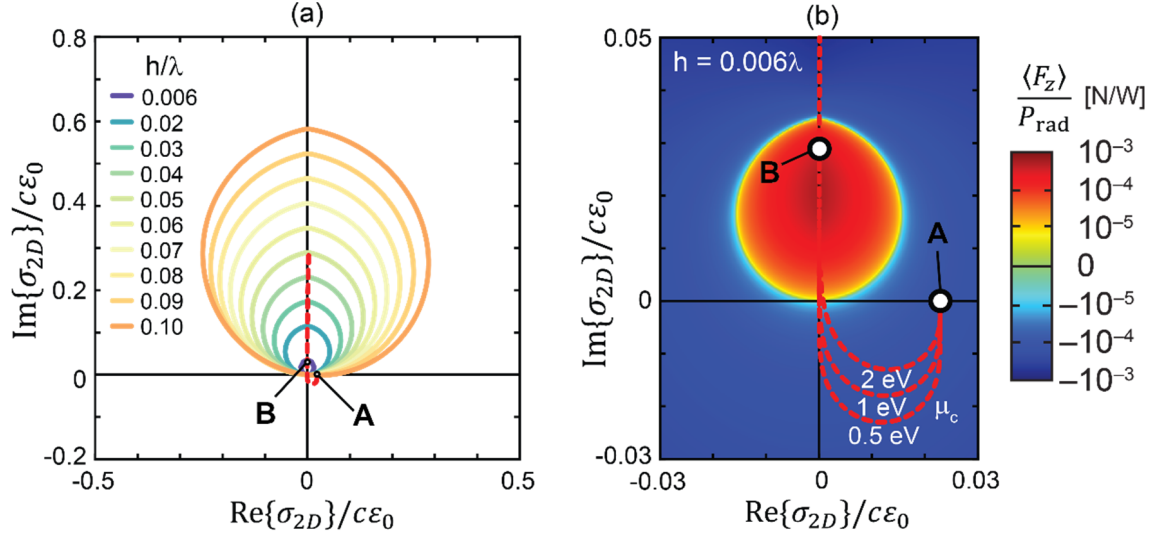

**Fig. 2.** (a) Contour plot in the complex plane of graphene conductivity $\sigma_{2D}$ enclosing the region of conductivity for which an attractive force takes place, for different dipole heights. (b) Time-averaged vertical force acting on the dipole plotted in the complex plane of $\sigma_{2D}$ [zoom in from (a)] for a fixed height h = 0.006λ. (Dashed red lines) parametric plot of the conductivity of graphene as the frequency is varied, for graphene with different chemical potentials obtained from the Kubo formula. The conductivity of ideal graphene is labelled "A", which coincides with the limit of Kubo formula at high frequencies $\hbar\omega \gg \mu_c$. The conductivity for graphene with a chemical potential $\mu_c$ at a certain frequency, arbitrarily chosen for strong repulsion in a region of metallic character, is labelled as "B". The conductivity of graphene for different values of $\mu_c$ crosses point "B" at different frequencies (e.g., for $\mu_c = 2$ eV, it happens at $\hbar\omega = 1.65$ eV). The plots correspond to a horizontal dipole $\mathbf{p} = p_x \hat{\mathbf{x}}$ over a free standing 2D material (taking $\varepsilon_1 = \varepsilon_3 = 1$).

**Optical forces on dipoles above graphene -** As a practical example of the ideas above, we will consider the behaviour of graphene. We would like to stress that all the above results are general and apply to any 2D material, and we only choose graphene as an example due to its well-characterized optical properties. The simplest ideal model for graphene is given by the conductivity $\sigma_{2D} = e^2/4\hbar$, which corresponds to $\sigma_{2D} \approx (0.023 + 0i)c\varepsilon_0$. This does not have the required imaginary part for repulsion, and we label this as case "A" in Figs 1-3. When a more realistic model of graphene with a non-zero chemical potential $\mu_c$ is used, its conductivity acquires a positive imaginary part at photon energies $\hbar\omega < (5/3)\mu_c$, responsible for its metallic character, and enabling repulsion. We model the conductivity of graphene $\sigma_{2D}(\omega)$ via the Kubo formula[22]–[24], see Supplementary Information. The wavelength-dependent conductivity of graphene with different chemical potentials $\mu_c$ are shown as red dashed lines on Fig. 2, from which the existence of conductivity values well inside the repulsion region appears clearly. We label an arbitrarily chosen point of the dashed line in this region as case "B" in Figs 1-3. This means that the chemical potential of graphene can be used to control and switch the repulsive force that it exerts on a nearby radiating dipole. The frequency-dependent force for a dipole above a graphene layer with chemical potential of $\mu_c = 0$ eV and 2 eV is shown in Figs. 3a-b. We see that the repulsion has a very broad bandwidth in the frequency region $0 < \hbar\omega < (5/3)\mu_c$. This observation is confirmed for other values of $\mu_c$, as shown in the Supplementary Information. Figure 3 also shows the associated electric field of the dipole near the graphene layer (Figs. 3(c-d)) and the time-averaged Poynting vector (Figs. 3(e-f)). We can see that the metallic character of graphene at these frequencies allows surface waves to be excited[25], and the power radiated from the dipole into the downward direction towards the surface couples into the surface waves (Fig. 3(d)). We hypothesise that there is a downward-directed flow of electromagnetic momentum, which must be accompanied by an upwards recoil mechanical force responsible for the repulsion, similar to the mechanism of propulsion in classical mechanics. When the dipole further approaches the surface, repulsion changes into attraction. This is seen in the potential energy landscape from Fig. 1(c) and in the force plot of Fig. 3(b).

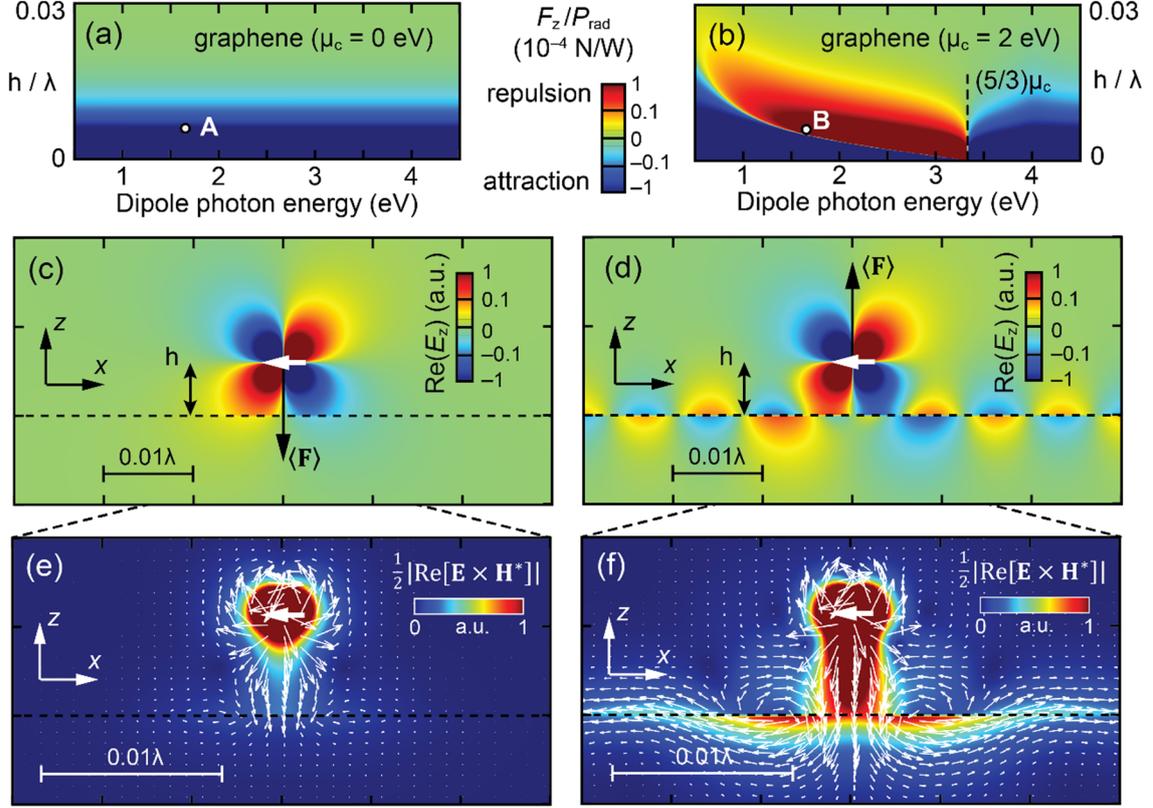

**Fig 3**. Optical force acting on a horizontal dipole over graphene as a function of dipole frequency $\hbar\omega$ and height of the dipole over the surface $h$, for graphene with a chemical potential (a) $\mu_c = 0$ eV (labelled "A" in Figs. 1,2) and (b) $\mu_c = 2$ eV (labelled "B" in Figs. 1,2). (c-d) Field plots ($E_z$) for the dipole over graphene as in (a,b), respectively, for a dipole height $h = 0.006\lambda$ and a dipole frequency $\hbar\omega = 1.65$ eV. (e-f) Associated time-averaged Pointing vector [zoom in (c,d)]. The fields were calculated by direct integration of the Green functions for the electric and magnetic fields, taken as in Ref.[26]. An exact match of the fields and forces to those from frequency-domain numerical simulations in CST Microwave Studio was conformed (see Supplementary Information).

**Forces on illuminated polarizable particles near two dimensional materials** - In the optical regime, a dipolar source is easily realized by the scattering of a small illuminated polarizable particle or molecule. In this case, in addition to the force caused by the dipole scattering itself, the illuminating light will also exert a gradient and scattering force on the particle. This situation can be analysed as follows: the particle gets polarized by all the fields incident on it, according to $\mathbf{p} = \alpha\left(\mathbf{E}_{pw}(\mathbf{r_0}) + \mathbf{E}_s^r(\mathbf{r_0})\right)$ where $\alpha$ is the isotropic polarizability, $\mathbf{E}_{pw}(\mathbf{r})$ is the superposition of the incident, reflected and transmitted plane waves, and $\mathbf{E}_s^r(\mathbf{r}) = \overleftrightarrow{\mathbf{G}}(\mathbf{r}, \mathbf{r_0}) \cdot \mathbf{p}$ is the back-scattering of the dipole fields reflected on the sheet. By solving this equation self-consistently for $\mathbf{p}$, we can compute the total time-averaged optical force acting on the particle as $\langle\mathbf{F}\rangle = (1/2)\sum_{i=x,y,z} Re\{p_i^* \nabla(E_{pw,i} + E_{s,i}^r)\}$. In this case, the source of energy is the plane wave, and the particle is just a passive scatterer.

This removes the infinities that appeared when the dipole was approaching the surface $h \to 0$. Fig. 4 shows the potential energy landscape for a polarizable particle with polarizability near a graphene layer corresponding to cases "A" ($\mu_c$= 0 eV) and "B" ($\mu_c$= 2 eV), when light is illuminated at normal incidence from below. It clearly shows the repulsion of the particle from the graphene layer at sub-wavelength distances. As in the dipole source case, at the smallest distances $h \to 0$ the force becomes attractive in both cases.

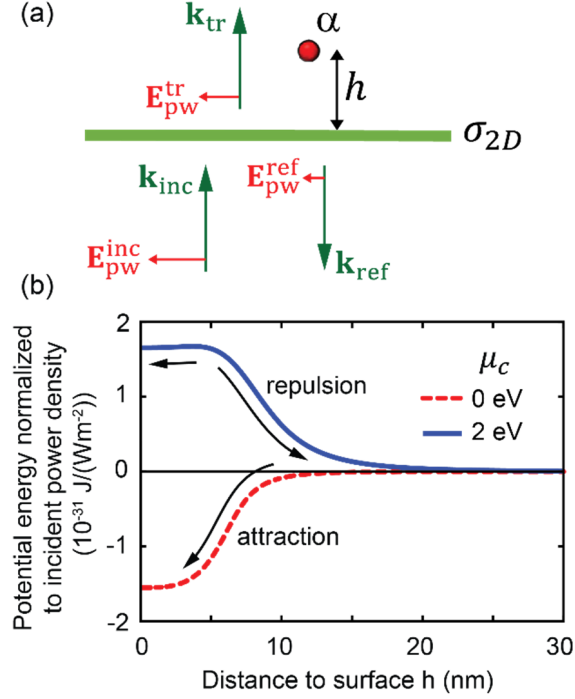

**Fig. 4** (a) Point-dipole-like particle near a graphene sheet illuminated under normal incidence from below (through the graphene sheet). (b) Potential energy landscape of a polarizable particle with polarizability $\alpha = (10^{-33})$ m$^3$ near graphene with chemical potential $\mu_c$ = 0 eV and 2 eV illuminated with light of frequency $\hbar\omega$ = 1.65 eV ($\lambda$ = 750 nm).

**Conclusions -** We have shown an extremely simple mechanism for the repulsion of small polarizable particles from conductive sheets such as graphene layers with an appropriate chemical potential. Our argument relies on the solution of a classical optics scattering problem assuming an ideal dipolar particle and a homogeneous conductivity model for two dimensional materials, therefore ignoring complexities such as higher order multipoles effects that may arise due to coupling with the surface, as well as the surface atomic arrangements, electronic band structures, and quantum effects. The fact that such a robust, strong and broadband repulsion arises from this simplest possible model suggests that the result is of fundamental nature, and we expect it to persist in more refined analysis. Note that repulsion near ENZ substrates was originally studied for dipoles[6], [7], but later found to be extensible to much more complex cases, such as finite-size antennas[27] and even optical repulsion of dielectric waveguides[28]. We expect the same extended applicability for repulsion from 2D materials.

We did not consider here other sources of forces on the particle, such as electrostatic charging, and fluctuation-induced forces (Casimir/London dispersion/van der Waals) caused by thermal and quantum fluctuations, which are known to dominate at small distances. In fluctuation electrodynamics, the computation of Casimir interactions can be reduced to solving the classical scattering problem, integrated over the frequency fluctuations[29]. This leads to an interesting possibility: since the classical scattering repulsion studied in this work exists on a wide frequency range, we expect that fluctuation-induced forces will be greatly affected. Thus, two dimensional sheets could have potential for applications in low friction devices[30] by exhibiting reduced attractive or perhaps even repulsive fluctuation-induced forces. This effect could be related with the experimentally observed unusual wetting characteristics of graphene on varying chemical potentials[11], [12].

**Acknowledgements:**

The authors acknowledge helpful discussions with N. Engheta. This work was supported by European Research Council project ERC-2016-STG-714151-PSINFONI and EPSRC (UK). A.Z. acknowledges support from the Royal Society and the Wolfson Foundation. All data supporting this research is provided in full in the text and supplementary information.

# Repulsion of polarized particles from two-dimensional materials

Francisco J. Rodríguez-Fortuño, Michela F. Picardi and Anatoly V. Zayats

*Department of Physics, King's College London, Strand, London WC2R 2LS, United Kingdom*

**SUPPLEMENTARY INFORMATION**

**Optical response of a two dimensional sheet**

The Fresnel coefficients of a two-dimensional sheet with sheet conductivity $\sigma_{2D}$ can be obtained by solving the electromagnetic boundary conditions $\hat{\mathbf{n}} \times (\mathbf{E}_1 - \mathbf{E}_3) = 0$ and $\hat{\mathbf{n}} \times (\mathbf{H}_1 - \mathbf{H}_3) = \mathbf{J}_s$, where $\hat{\mathbf{n}}$ is a unit vector perpendicular to the interface (in this case $\hat{\mathbf{n}} = \hat{\mathbf{z}}$), $\mathbf{E}_1, \mathbf{H}_1, \mathbf{E}_3, \mathbf{H}_1$ are the electric and magnetic fields in the upper and lower media, respectively, and $\mathbf{J}_s = \sigma_{2D}\mathbf{E}_t$ is the surface current density induced in the sheet, proportional to the tangential electric field $\mathbf{E}_t = \mathbf{E}_1 - \mathbf{E}_1 \cdot \hat{\mathbf{n}}$. Alternatively, two dimensional sheets with a sheet conductivity $\sigma_{2D}$ can also be modelled as shown in Supplementary Fig. S1, by considering a slab with a small thickness $\Delta \ll \lambda$, made up of a material with a thickness-dependent effective relative permittivity $\varepsilon_{\text{eff}}(\Delta) = 1 + i\sigma_{2D}/(\omega\varepsilon_0\Delta)$. The optical response of such a thin slab converges, in the limit $\Delta \to 0$, to the behaviour of the ideal two-dimensional material with conductivity $\sigma_{2D}$[7], [22]. Both procedures lead to the complex field transmission and reflection Fresnel coefficients of a two dimensional sheet sandwiched between materials with relative permittivities $\varepsilon_1$ and $\varepsilon_3$ given by:

$$r^{\text{p}}(k_{\text{t}}) = \frac{(\varepsilon_3+k_{z3r}\sigma_{\text{norm}})k_{z1r} - \varepsilon_1 k_{z3r}}{(\varepsilon_3+k_{z3r}\sigma_{\text{norm}})k_{z1r} + \varepsilon_1 k_{z3r}} \qquad t^{\text{p}}(k_{\text{t}}) = \frac{2(\varepsilon_1\varepsilon_3)^{1/2}k_{z1r}}{\varepsilon_3 k_{z1r} + \varepsilon_1 k_{z3r} + k_{z1r}k_{z3r}\sigma_{\text{norm}}}$$

$$r^{\text{s}}(k_{\text{t}}) = \frac{k_{z1r} - (k_{z3r} + \sigma_{\text{norm}})}{k_{z1r} + (k_{z3r} + \sigma_{\text{norm}})} \qquad t^{\text{s}}(k_{\text{t}}) = \frac{2k_{z1r}}{k_{z1r} + k_{z3r} + \sigma_{\text{norm}}} \tag{S1}$$

Where $k_{\text{tr}} = k_{\text{t}}/k_0 = (k_x^2 + k_y^2)/k_0$ is the normalized wave-vector component in the plane parallel to the surface (conserved at the interfaces), $k_{zir} = k_{zi}/k_0 = (\varepsilon_i - k_{\text{tr}}^2)^{1/2}$ is the normalized wave-vector component in the direction perpendicular to the sheet at the $i$-th medium, and $\sigma_{\text{norm}} = \sigma_{2D}/(c\varepsilon_0)$ is a dimensionless way to express the two dimensional sheet conductivity. The normalization of conductivity and wave-vectors allows Eqs. S1 to be written compactly, and all the quantities involved are dimensionless. The expressions are simplified even further if the substrate and superstrate are equal ($\varepsilon_1 = \varepsilon_3$): for example, $t^p_{\varepsilon_1=\varepsilon_3} = 2\varepsilon_1/(2\varepsilon_1 + k_{zr}\sigma_{\text{norm}})$. Eqs. (S1) are valid for both propagating ($|k_t| < k_1$) and evanescent ($|k_t| \geq k_1$) components, and therefore can be used to calculate the fields resulting from any incident field of which its spatial Fourier decomposition $E^{p/s}(k_x, k_y)$ is known, including the fields of a dipole source.

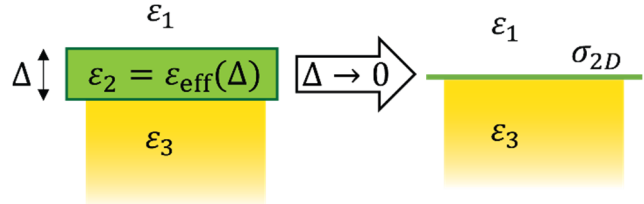

**Fig S1**. A two dimensional sheet with conductivity $\sigma_{2D}$ can be modelled as a thin slab of thickness $\Delta$ with thickness-dependent permittivity $\varepsilon_{\text{eff}}(\Delta) = 1 + i\sigma_{2D}/(\omega\varepsilon_0\Delta)$, whose Fresnel coefficients converge in the limit $\Delta \to 0$.

## Green function's gradient

The fields reflected by a dipole located at $\mathbf{r}_0$ near a surface defined by $z = 0$ are given in terms of the Green function as $\mathbf{E}(\mathbf{r}, \omega) = \overleftrightarrow{\mathbf{G}}(\mathbf{r}, \mathbf{r_0}, \omega)\,\mathbf{p}$. The reflected field Green function for a surface with arbitrary reflection coefficients $r^p(k_t)$ and $r^s(k_t)$ can be written using Weyl's identity[16], [17] as:

$$\overleftrightarrow{\mathbf{G}}(\mathbf{r}, \mathbf{r}_0, \omega) = \frac{i}{8\pi^2 \varepsilon_0 \varepsilon_1} \iint dk_x dk_y e^{i(k_x(x-x_0)+k_y(y-y_0)+k_{z1}(z+z_0))} \left[r^p \overleftrightarrow{\mathbf{M}}_p + r^s \overleftrightarrow{\mathbf{M}}_s\right] \quad (S2)$$

where the integral is performed over $k_x, k_y \in [-\infty, \infty]$ and $\overleftrightarrow{\mathbf{M}}_p$ and $\overleftrightarrow{\mathbf{M}}_s$ represent the p- and s- polarized components of the dipole, given by:

$$\overleftrightarrow{\mathbf{M}}_p = \begin{pmatrix} -k_{z1} k_x^2/k_t^2 & -k_{z1} k_x k_y/k_t^2 & -k_x \\ -k_{z1} k_x k_y/k_t^2 & -k_{z1} k_y^2/k_t^2 & -k_y \\ k_x & k_y & k_t^2/k_{z1} \end{pmatrix}$$

$$\overleftrightarrow{\mathbf{M}}_s = \frac{k_1^2}{k_{z1} k_t^2} \begin{pmatrix} k_y^2 & -k_x k_y & 0 \\ -k_x k_y & k_y^2 & 0 \\ 0 & 0 & 0 \end{pmatrix}$$

with $k_1 = n_1 k_0 = n_1 \omega/c$. The gradient of the Green function is performed with respect to $\mathbf{r} = (x, y, z)$ and is given by:

$$\nabla \overleftrightarrow{\mathbf{G}} = \frac{\partial \overleftrightarrow{\mathbf{G}}}{\partial x}\hat{\mathbf{x}} + \frac{\partial \overleftrightarrow{\mathbf{G}}}{\partial y}\hat{\mathbf{y}} + \frac{\partial \overleftrightarrow{\mathbf{G}}}{\partial z}\hat{\mathbf{z}}$$

where the spatial derivatives can be directly obtained from Eq. (S2) as:

$$\frac{\partial \overleftrightarrow{\mathbf{G}}}{\partial x}(\mathbf{r}, \mathbf{r}_0, \omega) = -\frac{1}{8\pi^2 \varepsilon_0 \varepsilon_1} \iint k_x dk_x dk_y e^{i(k_x(x-x_0)+k_y(y-y_0)+k_{z1}(z+z_0))} \left[r^p \overleftrightarrow{\mathbf{M}}_p + r^s \overleftrightarrow{\mathbf{M}}_s\right]$$

$$\frac{\partial \overleftrightarrow{\mathbf{G}}}{\partial y}(\mathbf{r}, \mathbf{r}_0, \omega) = -\frac{1}{8\pi^2 \varepsilon_0 \varepsilon_1} \iint k_y dk_x dk_y e^{i(k_x(x-x_0)+k_y(y-y_0)+k_{z1}(z+z_0))} \left[r^p \overleftrightarrow{\mathbf{M}}_p + r^s \overleftrightarrow{\mathbf{M}}_s\right]$$

$$\frac{\partial \overleftrightarrow{\mathbf{G}}}{\partial z}(\mathbf{r}, \mathbf{r}_0, \omega) = -\frac{1}{8\pi^2 \varepsilon_0 \varepsilon_1} \iint k_{z1} dk_x dk_y\, e^{i(k_x(x-x_0)+k_y(y-y_0)+k_{z1}(z+z_0))} \left[r^p \overleftrightarrow{\mathbf{M}}_p + r^s \overleftrightarrow{\mathbf{M}}_s\right]$$

The gradient needs to be calculated at the location of the dipole, therefore we take the limit $\mathbf{r} \to \mathbf{r}_0$ and we assume $\mathbf{r}_0 = (0,0,h)$ which simplifies the expressions to:

$$\frac{\partial \overleftrightarrow{\mathbf{G}}}{\partial x}(\mathbf{r}_0, \mathbf{r}_0, \omega) = -\frac{1}{8\pi^2 \varepsilon_0 \varepsilon_1} \iint k_x dk_x dk_y e^{i2k_{z1}h} \left[r^p \overleftrightarrow{\mathbf{M}}_p + r^s \overleftrightarrow{\mathbf{M}}_s\right]$$

$$\frac{\partial \overleftrightarrow{\mathbf{G}}}{\partial y}(\mathbf{r}_0, \mathbf{r}_0, \omega) = -\frac{1}{8\pi^2 \varepsilon_0 \varepsilon_1} \iint k_y dk_x dk_y e^{i2k_{z1}h} \left[r^p \overleftrightarrow{\mathbf{M}}_p + r^s \overleftrightarrow{\mathbf{M}}_s\right]$$

$$\frac{\partial \overleftrightarrow{\mathbf{G}}}{\partial z}(\mathbf{r}_0, \mathbf{r}_0, \omega) = -\frac{1}{8\pi^2 \varepsilon_0 \varepsilon_1} \iint k_{z1} dk_x dk_y e^{i2k_{z1}h} \left[r^p \overleftrightarrow{\mathbf{M}}_p + r^s \overleftrightarrow{\mathbf{M}}_s\right]$$

We can now write the transverse wave-vectors in cylindrical coordinates $k_x = k_t \cos\alpha$ and $k_y = k_t \sin\alpha$ and, with some algebra, perform the angular integration in $\alpha \in [0, 2\pi]$ leaving only a single integration in $k_t \in [0, \infty]$:

$$\frac{\partial \overleftrightarrow{\mathbf{G}}}{\partial x}(\mathbf{r}_0, \mathbf{r}_0, \omega) = \frac{1}{8\pi\varepsilon_0\varepsilon_1} \int dk_t e^{i2k_{z1}h} k_t^3 r^p \begin{pmatrix} 0 & 0 & 1 \\ 0 & 0 & 0 \\ -1 & 0 & 0 \end{pmatrix}$$

$$\frac{\partial \overleftrightarrow{\mathbf{G}}}{\partial y}(\mathbf{r}_0, \mathbf{r}_0, \omega) = \frac{1}{8\pi\varepsilon_0\varepsilon_1} \int dk_t e^{i2k_{z1}h} k_t^3 r^p \begin{pmatrix} 0 & 0 & 0 \\ 0 & 0 & 1 \\ 0 & -1 & 0 \end{pmatrix}$$

$$\frac{\partial \overleftrightarrow{\mathbf{G}}}{\partial z}(\mathbf{r}_0, \mathbf{r}_0, \omega) = \frac{1}{8\pi\varepsilon_0\varepsilon_1} \int dk_t e^{i2k_{z1}h} k_t \begin{pmatrix} k_{z1}^2 r^p - k_1^2 r^s & 0 & 0 \\ 0 & k_{z1}^2 r^p - k_1^2 r^s & 0 \\ 0 & 0 & -2k_t^2 r^p \end{pmatrix}$$

The time-averaged force vector acting on the dipole is given as:

$$\langle \mathbf{F} \rangle = \sum_{i,j=x,y,z} \frac{1}{2} Re\{p_i^* p_j \nabla G_{ij}\} = \sum_{i,j=x,y,z} \frac{1}{2} Re\left\{ p_i^* p_j \left( \frac{\partial G_{ij}}{\partial x} \hat{\mathbf{x}} + \frac{\partial G_{ij}}{\partial y} \hat{\mathbf{y}} + \frac{\partial G_{ij}}{\partial z} \hat{\mathbf{z}} \right) \right\}$$

Where the sum is done over nine terms, corresponding to the nine elements of the tensor, most of which are zero. The z-component of the force is therefore given by:

$$\langle \mathbf{F} \rangle \cdot \hat{\mathbf{z}} = \frac{1}{2} Re\left\{ |p_x|^2 \frac{\partial G_{xx}}{\partial z} + |p_y|^2 \frac{\partial G_{yy}}{\partial z} + |p_z|^2 \frac{\partial G_{zz}}{\partial z} \right\}$$

which after substitution of the appropriate terms results in the equation given in the main text. Notice that if the dipole is linearly polarized, we have $p_i^* p_j = p_j^* p_i$, and the lateral components of the force exactly cancel out.

**Kubo formula for the conductivity of graphene**

In this work we used the Kubo formula[22]–[24] to model the conductivity of graphene, which is known to fit well with experimental results. The temperature was taken as T = 293 K, the energy gap as 0 eV, and the scattering rate was set at $\Gamma$=1.29 meV, which is the highest amongst Refs.[22]–[24], and the results do not depend strongly on the scattering rate.

**Effect of dipole polarization and presence of a substrate**

Supplementary Figure 2 shows the frequency dependent vertical force acting on a vertical dipole above graphene at different chemical potentials, comparing free standing graphene ($\varepsilon_3 = 1$) with the case of a substrate with $\varepsilon_3 = 2$. By comparison with the figures in the main text, it is clearly seen that the behaviours discussed in the main text are robust to changes in dipole polarization. The presence of the substrate has an interesting effect in the region of repulsive force (Fig S2(b,d,f)). Clearly, as the substrate's refractive index increases, the repulsive region will get smaller and eventually disappear, but the repulsion effects exists for low index substrates. Interestingly, for the case $\varepsilon_3 > 1$ we see that the near field force as a function of distance has two sign changes for a given frequency. This implies a stable point of equilibrium in the potential energy landscape. The equilibrium height depends on the frequency, which suggests interesting applications for particle sorting and manipulation.

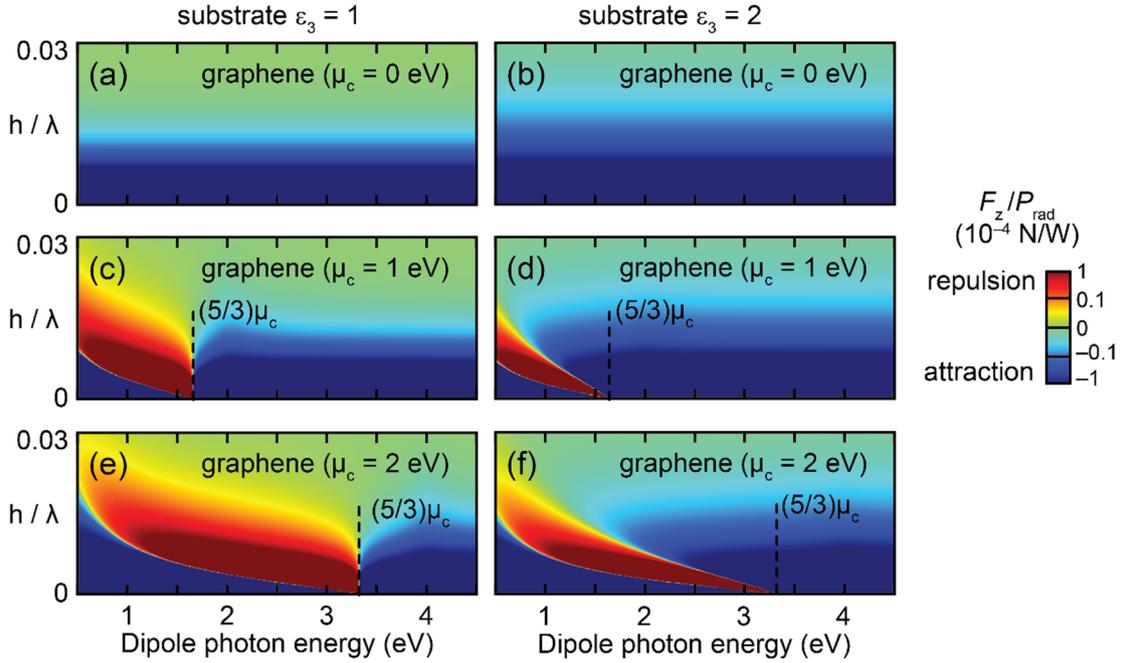

**Fig. S2.** Map of the vertical force for a vertical dipole over a graphene sheet with different chemical potentials and for different substrate permittivities. The observed behaviour is qualitatively identical to a horizontal dipole considered in the main text. The upper frequency limit of the repulsion can be seen to converge to $(5/3)\mu_c$.

**Numerical simulations**

Supplementary Figure 3 shows the comparison between the forces calculated analytically using Eq. (1) and numerically by integrating Maxwell's stress tensor using the commercial software CST Microwave Studio. Both theory and simulations are in excellent agreement.

In the numerical simulations, infinitely thin two-dimensional sheets cannot be modelled. Instead, we modelled graphene as a very thin film of thickness $\Delta = 0.5$ nm with corresponding permittivity $\varepsilon_{\text{eff}}(\Delta)$ as detailed above. In the figure we also show the analytical calculation of the force for the case of that same thin sheet, seen to give results very close to the infinitely thin sheet.

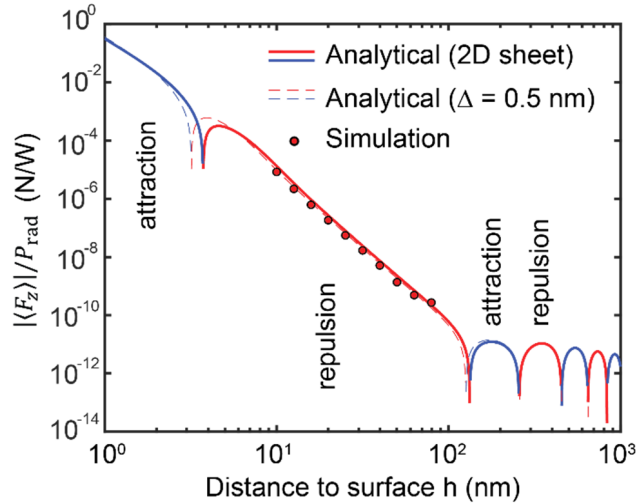

**Fig. S3**. Logarithmic plot of the distance dependence of the vertical force for a horizontal dipole over a graphene sheet, corresponding to case "B" (frequency $\hbar\omega=1.65$ eV and graphene chemical potential $\mu_c = 2$ eV). Red and blue curves correspond to repulsive and attractive forces, respectively. Comparison between the forces calculated using the numerical integration of Eq. (1) in the main text for graphene modelled as an infinitely thin layer (analytical limit) and as a 0.5 nm thick layer, and using simulations with CST Microwave Studio in which graphene is modelled as a 0.5 nm thick layer. In the simulations, the force was obtained from the fields by integration of Maxwell's stress tensor. The force was checked to be independent of the size of the integration volume around the dipole, showing robustness of the result. Numerical noise was relatively large in the last data point due to the very low value of the force.

13